%% file: ms_final.tex
\documentclass[structabstract]{aa}
\usepackage{graphicx}
\usepackage{amsmath}
\usepackage{amssymb}
\usepackage{epsfig} 
\usepackage{txfonts}
\usepackage{color}
\usepackage{xcolor}
\usepackage{url}
\usepackage{natbib}
\usepackage{soul}
\bibpunct[]{(}{)}{;}{a}{}{,}

\newcommand{\K}{\rm\thinspace K}
\newcommand{\Msun}{\hbox{$\rm\thinspace \text{M}_{\odot}$}}

\begin{document}

\title{Asteroseismic inversions for radial differential rotation of Sun-like stars: ensemble  fits}
\titlerunning{Asteroseismic ensemble fitting for radial differential rotation}

\author{
H.~Schunker \inst{\ref{inst1}}
\and
J.~Schou \inst{\ref{inst1}}
\and
W.~H.~Ball \inst{\ref{inst2}}
\and
M.~B.~Nielsen  \inst{\ref{inst2}}
\and
L.~Gizon  \inst{\ref{inst1},\ref{inst2}}
}

\institute{
Max-Planck-Institut f\"{u}r Sonnensystemforschung, Justus-von-Liebig-Weg 3, 37077  G\"{o}ttingen, Germany \label{inst1}\\
\email{schunker@mps.mpg.de}
\and
Georg-August-Universit\"{a}t G\"{o}ttingen, Institut f\"{u}r Astrophysik, Friedrich-Hund-Platz 1, 37077   G\"{o}ttingen, Germany\label{inst2}\\
}

\date{Received $\langle$date$\rangle$ / Accepted $\langle$date$\rangle$}

\abstract
%context 
{
Radial differential rotation is an important parameter for stellar dynamo theory and for understanding angular momentum transport.
}
%aim
{
We investigate the potential of using a large number of similar stars simultaneously to constrain their average radial differential rotation gradient: we call this `ensemble fitting'.
}
%method
{
We use a range of stellar models along the main sequence, each with a synthetic rotation profile. 
The rotation profiles are step functions with a step of $\Delta \Omega = -0.35\mu$Hz, which is located at the base of the convection zone. 
These models are used to compute the rotational splittings of the p modes and to model their uncertainties. 
We then fit an ensemble of stars to infer the average $\Delta \Omega$.
}
%results
{
All the uncertainties on the inferred $\Delta \Omega$ for individual stars are  of the order $1\mu$Hz. 
Using 15 stellar models in an ensemble fit, we show that the uncertainty on the average $\Delta \Omega$ is reduced to less than the input $\Delta \Omega$, which allows us to constrain the sign of the radial differential rotation. 
We show that a solar-like $\Delta \Omega \approx 30$~nHz can be constrained by an ensemble fit of thousands of main-sequence stars.
Observing the number of stars required to successfully exploit the ensemble fitting method will be possible with future asteroseismology missions, such as PLATO.
We demonstrate the potential of ensemble fitting by showing that any systematic differences in the average $\Delta \Omega$ between F, G, and K-type stars larger than $100$~nHz can be detected. 
}
{}
\maketitle

%\tableofcontents 

%%%%%%%%%%%%%%%%%%%%%%

\section{Introduction}\label{sect:intro}

One of the features of many solar dynamo models is the rotational shear layer below the convection zone, the tachocline \citep{SpiegelZahn1992}, where the magnetic field is thought to be generated in some models \cite[e.g.][]{Charbonneau2010}. Despite knowing the spatially-resolved internal rotation profile of the Sun \citep{Schouetal1998}, it is poorly constrained in other Sun-like stars. 
Asteroseismology is the only tool available to  measure the  rotation of stellar interiors. 

In the power spectrum, modes of oscillation have strong peaks at their oscillation frequency.
When the rotation rate is slow the oscillation is perturbed so that the mode becomes a multiplet in frequency,
$\omega_{nlm} = \omega_{n\ell} + m \delta \omega_{n\ell}$, ignoring any latitudinal differential rotation, where $n$ is the radial order, $\ell$ is the harmonic degree, and $m$ is the azimuthal order of the mode, described by a spherical harmonic. These azimuthal components are called splittings.

From the splittings, an inversion technique can be used to infer the internal rotation profile.
This has been performed for stars with large rotation rates (and easily measured splittings) and modes sensitive to different depths, tightly constraining the core rotation rate \citep[e.g.][]{Deheuvelsetal2012,Deheuvelsetal2014}.
However, in Sun-like stars, inferring the radial differential rotation (RDR, hereafter)  is more challenging. 
\citet{Lundetal2014b} found that it is unlikely that asteroseismology of Sun-like stars will result in reliable inferences of the RDR profile, such as can be performed for red giants.
This is predominantly due to the degeneracy of the information contained in the observable modes: they are sensitive to roughly the same regions of the stellar interior \citep[see Figs.~3 and 4 in ][ Erratum]{Lundetal2014b}.
They posit that, at best, it may be possible to obtain the sign of the RDR gradient.
Despite the inherent challenges, rotational splitting as a function of frequency has been measured in Sun-like stars  by \citet{Nielsenetal2014,Nielsenetal2015}  using the short-cadence long duration observations from \textit{Kepler} \citep{Boruckietal2010}. 

\citet[][, hereafter Paper~I]{Schunkeretal2015a} conclude that it would not be possible to infer a radially resolved rotation profile for a single Sun-like star using regularised inversion techniques. 
As an alternative, they demonstrate that direct functional fitting of a step-function can constrain the sign of the RDR gradient better and  can even be used to retrieve the surface rotation rate of a subgiant star.

Paper~I  also explores the sensitivity of linear inversions to the stellar models used in the inversion. They conclude that the inversions are insensitive to the uncertainties in the stellar model, compared to the level of noise in the splittings. They therefore propose that it may be possible to exploit this insensitivity to perform ensemble asteroseismic inversions across a broader range of stellar types.

Ensemble asteroseismic inversions have the advantage that the uncertainties will be reduced with the extra information from more stars, from which an ensemble inversion can be solved to infer an average RDR profile. 
The upcoming PLATO~2.0 mission \citep{Rauer2013} will observe tens of thousands of solar-like stars, offering us the opportunity to exploit this ensemble fitting method to constrain the  RDR gradient in Sun-like stars. 

In this paper we explore the possibility of using ensemble asteroseismic fits of Sun-like stars to infer the average RDR.
We first  solve the forward problem to compute the rotational splittings for a broad range of stellar models with a synthetic rotation profile (Sect.~\ref{sect:forward}). 
We describe the fitting function and the ensemble fit method (Sect.~\ref{sect:funcfit}). 
We show the results for independent fits (Sect.~\ref{sect:indfit}) and apply the ensemble fit to all 15 stellar models in Sect.~\ref{sect:ens}. 
We then extend our results to estimate how well we can constrain the size of the step function along the main sequence using subsets of the Sun-like stars that are expected to be observed by PLATO  (Sect.~\ref{sect:plato}).
We present our conclusions in Sect.~\ref{sect:conc}.

\section{The forward problem: modelling splittings}\label{sect:forward}

%%%%%%%%%%%%%%%%%
\subsection{Stellar models}\label{sect:models}

%Stellar models
We compute a set of 15 stellar models of main sequence solar-like oscillators that covers F to K-type stars, with a range of metallicities and ages using the Modules for Experiments in Stellar Astrophysics code \citep[MESA\footnote{\url{http://mesa.sourceforge.net/}} revision 6022][]{Paxton2013}. %{paxton2011,paxton2013,paxton2015}.
Opacities are taken from OPAL \citep{Iglesias1996} and \citet{Ferguson2005} at high and low temperatures, respectively.  The equation of state tables are based on the 2005 version of the OPAL tables \citep{Rogers2002}, and nuclear reaction rates are drawn from the NACRE compilation \citep{Angulo1999} or \citet{Caughlan1988}, if not available in the NACRE tables.  We used newer specific rates used for the reactions
${}^{14}\mathrm{N}(p,\gamma){}^{15}\mathrm{O}$ \citep{Imbriani2005}
and ${}^{12}\mathrm{C}(\alpha,\gamma){}^{16}\mathrm{O}$ \citep{Kunz2002}. Convection was formulated using standard mixing-length theory \citep{Boehm-vitense1958} with mixing-length parameter $\alpha=2$.
The solar metallicity and mixture was that of \citet{Grevesse1998} and atomic diffusion was included, using the prescription of \citet{Thoul1994}.

Models were initialised on the pre-main-sequence  with a central temperature $T_c=9\times10^5\K$.  We computed tracks for masses $0.8$, $1.1$ and $1.4\Msun$ and initial metallicities $[\mathrm{Fe/H}]=-0.5$ and $0.5$.  Along each track, we recorded the models with central hydrogen abundances $X_c=0.60$, $0.35,$ and $0.10$ to represent stars towards the beginnings, middles, and ends of their main-sequence lives.  We excluded models from the low-metallicity $1.4\Msun$ track because all had effective temperatures above $7000\K$ and would therefore either not be solar-like oscillators, or their oscillation frequencies would be difficult to identify and measure. The parameters of the models are presented in Table~\ref{table:ensmodels}. The rotational kernels for each mode of oscillation in the star can be computed from these models.
%%%   TABLE OF MODELS HERE   %%%
\begin{table}
\caption{Parameters of the stellar models used in the ensemble asteroseismic fits.}  % title of Table
\vspace{-0.5cm}
\label{table:ensmodels} % is used to refer this table in the text
\centering % used for centering table
\include{ense_models_table}
%\tablefoot{some comments here}
\end{table}

%%%%%%%%%%%%%%%%%%
\subsection{Synthetic rotation and mode splittings}\label{sect:rotation}
%Synthetic rotation profile
The Sun's convection zone shows radial and latitudinal differential rotation. The analysis of the acoustic modes which have low sensitivity in the deep interior suggests solid body rotation in the core \citep{Schouetal1998}.
Thus, we chose a step-function profile as the simplest representation of the expected rotation profile of a star.
Specifically, we chose\begin{equation}
        \Omega(r) =\left\{ 
                \begin{array}{l l l}
                \Omega_\mathrm{c}&=&1035~\mathrm{nHz} \quad 0 < r/R \le r_\mathrm{BCZ} \\    
                \Omega_\mathrm{c} + \Delta \Omega_\mathrm{c}&=&685~\mathrm{nHz} \quad r_\mathrm{BCZ} < r/R \le R
%               \Omega_\mathrm{e}=685~\mathrm{nHz} \quad r_\mathrm{BCZ} < r/R \le 1.0        
        \end{array} \right.     
,\end{equation} 
where $\Delta \Omega_\mathrm{c}=-350$~nHz, $R$ is the stellar radius, and $r_\mathrm{BCZ}$ is the radius of the base of the convection zone for the particular stellar model (see Table~\ref{table:ensmodels}). 

We ignore any latitudinal differential rotation.
The rotational splittings are a linear function of the angular rotation rate, $\delta \omega_{n\ell } = \int_0^R K_{n\ell}(r)\Omega(r) \mathrm{d}r $, where $K_{n\ell}$ are the rotational kernels of the star. For $\delta \omega_{n2}$ we compute the error-weighted mean of the splittings for the $|m|=1,2$ components, where the errors are the uncertainties modelled in  Sect.~\ref{sect:noise}.
%When the rotation rate is slow the oscillation is perturbed so that the mode becomes a multiplet  in frequency,
%$\omega_{nlm} = \omega_{n\ell} + m \delta \omega_{n\ell}$, ignoring any latitudinal differential rotation, where $n$ is the radial order, $\ell$ is the harmonic degree, and $m$ is the azimuthal order of the mode described by a spherical harmonic.

%%%%%%%%%%%%%%%%%%
\subsection{Noise model for the rotational splittings}\label{sect:noise}

Global rotational splittings have been unambiguously measured  \citep[e.g.][]{Gizonetal2013} for Sun-like stars.
The only attempt to measure variations in the splittings for Sun-like solar-like oscillators was done as a function of mode-set  \citep[$\ell=2,0,1$, ][]{Nielsenetal2014},  and not for independent modes.
Therefore, we need to model the measured uncertainties on the splittings.
We scale the uncertainties on the splittings from a function of the uncertainties on the frequencies, which is similar to Paper~I.

We begin with a model for the uncertainties on frequencies from  \citet[][, Eqn.~2]{Libbrecht1992}:
\begin{equation}
\sigma^2(\omega_{\ell m}) = f(\beta) \frac{\Gamma}{4 \pi T}
\label{eqn:frequnc} 
,\end{equation}
where $\Gamma$  is the FWHM of the mode linewidth, $T$ is the observing time, which we set to the expected three years of future asteroseismology observing campaigns, and $f(\beta)=(1+\beta)^{1/2}[(1+\beta)^{1/2} + \beta^{1/2}]^3$, where $\beta$ is the inverse signal-to-noise ratio $B(\nu)/H_{\ell m}$. 
Here we are specifically employing  $\nu$ to indicate a function of frequency, to distinguish against our use of $\sigma(\omega_{\ell m}),$ which indicates the uncertainty in the mode frequency  $\omega_{\ell m}$.
Here, $B(\nu)$ is the background noise spectrum and $H_{\ell m}$ is the mode height. We determine $\Gamma$ relative to the radial order of the mode closest to the frequency of maximum power, $\nu_\mathrm{max}$. $H_{\ell m}$ and $B(\nu)$ are scaled from the solar values relative to $\nu_\mathrm{max}$.

We first model the linewidths of the Sun by fitting the values for each mode in Table~2.4 of \cite{Stahn2011} with a third order polynomial as a function of radial order relative to the mode closest to  $\nu_\mathrm{max}$ to get $\Gamma(n)$. 
The function is then scaled using the relation $\Gamma/\Gamma_\odot \propto T_\mathrm{eff}^4/T_{\odot \, \mathrm{eff}}^4$ \citep{Chaplinetal2009} for each star.

We then model the envelope of mode heights in the Sun using the function
\begin{equation}
H_{\ell m}(\nu)=\frac{\mathcal{A}_\ell^2}{\pi \Gamma} \mathcal{E}_{\ell m}(i) P(\nu,\nu_c,\sigma_1,\sigma_2)
\end{equation}
\citep{Stahn2011}  relative to $\nu_\mathrm{max}$, and scaled by the surface gravity $H(\nu)/H_\odot(\nu) \propto g_\odot^2/g^2$  \citep{Chaplinetal2009}. 
The parameters of the amplitude $\mathcal{A}_\ell$, and $\nu_c$, $\sigma_1$, $\sigma_2$ of the asymmetric Lorentzian, $P$, are given in \citet[Table~2.3, ][]{Stahn2011}.
Here, we choose the inclination of the stellar rotation axis to be $i=90^\circ$ in the mode visibility term, $\mathcal{E}_{\ell m}(i)$ since, statistically, this inclination is most likely. We discuss the statistical effect of random inclinations later  in this paper (Sect.~\ref{sect:plato}).
The distribution of multiplet power as a function of stellar inclination for $\ell=1,2$ modes can be found in \citet{GizonSolanki2004}.

The background term, $B(\nu)$, is modelled using two Harvey laws  \citep{Harvey1985} with solar values from Table~2.1 of \citet{Stahn2011}, which are scaled by the frequency of maximum power, $B/B_\odot \propto \nu_\mathrm{max}/\nu_{\odot \, \mathrm{max}}$.
We note that including the background term increases the uncertainties in the splittings by up to 20\% for the stars with higher $T_\mathrm{eff}$.

%Mode lifetimes are inversely proportional to the effective temperature $\langle \frac{1}{\Gamma} \rangle  \propto T_\mathrm{eff}^{-4}$  and the height of the mode peaks in the intensity power spectrum (which determines the signal-to-noise - not the amplitude), scales with the surface gravity, $g^{-2}$  \citep{Chaplinetal2009}.
%\citet{CDF1983} determined that the p-mode velocity amplitude scales with the ratio of luminosity to mass, $\nu_\mathrm{osc} \propto \frac{L}{M}$. 
% Using this and the scaling relation $\frac{\delta L}{L} \propto \frac{\nu_\mathrm{osc}}{T^2_\mathrm{eff}}$ from \citet{KB1995}, the height of the intensity is proportional to the amplitudes and the linewidth, $H \propto \frac{A^2}{\Gamma} \propto g^{-2}$.
%Therefore,  $\Gamma \propto T^4_\mathrm{eff}$ and $A^2 \propto \frac{T^4_\mathrm{eff}}{g^2}$.
% $A^2 \propto H\Gamma \propto \frac{\Gamma}{g^2} \propto \frac{T^4_\mathrm{eff}}{g^2}$ (since $\Gamma \propto T^4_\mathrm{eff}$).

\begin{center}
\begin{figure}
\includegraphics[width=0.45\textwidth]{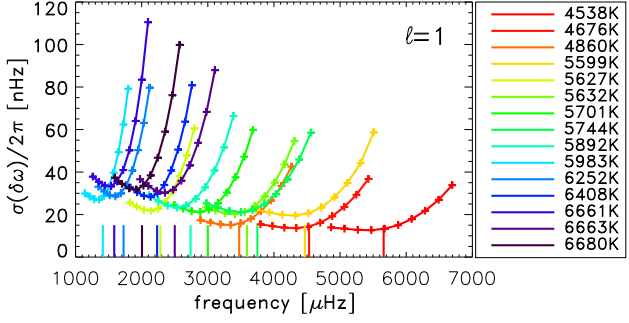} 
%   plot_ensemble_noise.pro 
\caption{
Uncertainty on the mode splittings $\ell=1$, $|m|=1$  for each stellar model, as indicated by the temperature. 
The vertical ticks indicate the frequency of maximum p-mode power, $\nu_\mathrm{max}$, in the envelope.
Hotter stars have modes with higher oscillation frequencies and larger uncertainties on the splittings.
}
\label{fig:ensnoise}
\end{figure}
\end{center}

Lastly, we compute the uncertainties for the frequency splittings slightly differently to Paper~I, which assumed that all modes were equally visible. 
The uncertainty on the splittings is then $\sigma_{m}(\delta \omega) = \sigma_{m}(\omega)/m$, and the average for each harmonic degree is $1/\sigma_\ell(\delta \omega)^2 = \sum_{m=-\ell}^{\ell} m^2/\sigma_{m}(\omega)^2$.
%For $i=90^\circ this is, $\sigma_{\ell=1}(\delta \omega) = \sigma_{\ell=1}(\omega)/\sqrt{2}$ and $\sigma_{\ell=2}(\delta \omega) = \sigma_{\ell=2}(\omega)/2\sqrt{2}$ 
Our final model for the uncertainties on the splittings is a quadratic  fit to the uncertainties of the splittings for each $\ell$ relative to $\nu_\mathrm{max}$.
Figure~\ref{fig:ensnoise} shows the uncertainties for the $\ell=1$ modes at an inclination of $90^\circ$ with  ten~radial orders evenly distributed about the frequency of maximum power. The uncertainties on the $\ell=2$ modes are half that of the $\ell=1$ modes.
We add a Gaussian-distributed random realisation of the noise that is centred at zero and with standard deviation given by the uncertainties to the splittings of each stellar model.
%\begin{equation}
%\sigma_{\ell m}(\delta \omega)=\sigma_{\ell m}(\omega)/\sqrt{\frac{1}{3} \ell(\ell+1)(2\ell+1)}.
%\end{equation}
%This relationship assumes that the uncertainties are independent, which could be improved upon in the future. 

%%%%%%%%%%%%%%%%%%
 \section{Inversion method: functional fitting}\label{sect:funcfit}
We call forcing a specific function to be fit to the data `functional fitting'.
We choose to fit a step function  as the simplest estimate of the rotation above and below the base of the convection zone, where the size of the step is a proxy for the globally averaged RDR gradient, from the core to the surface. Traditionally one would fit for the rotation rate below and above the convection zone.
We reformulate this slightly so that we fit for the rotation rate below the convection zone and for the size of the step. For a single star the function is 
\begin{equation}
F(\mathrm{r}) =\left\{ 
     \begin{array}{ll}
         \Omega^* &  0 < \mathrm{r/R} \le \mathrm{r}_\mathrm{BCZ}\\     
         \Omega^* + \Delta \Omega & \mathrm{r}_\mathrm{BCZ} < \mathrm{r/R} \le 1,
     \end{array} \right.
\end{equation}
where we fit the rotation rate below the base of the convection zone, $\Omega^*$, and the step size for the coefficients, $\Delta \Omega$.
In this way we can directly get a measure of the overall gradient of the RDR and its uncertainties.
The rotational splitting is linearly related to the coefficients $\delta \omega_i = \sum_{k=1}^2 c_{ik} a_k$, where $a_1=\Omega^*$ and $a_2=\Delta \Omega$,  $c_{i1}=\int_0^R K_{i}(r) \mathrm{d}r$ and $c_{i2}=\int_{\mathrm{r}_\mathrm{BCZ}}^R K_{i}(r) \mathrm{d}r$.

For an ensemble fit to $N$ stars, we fit for independent rotation rates below the convection zone for each star, $\Omega^*_N$, but for a common step size for all stars, $\Delta \Omega$. The splitting is  then the sum over all coefficients, $\delta \omega_i = \sum_{k=1}^{N+1} c_{ik} a_k$, where $a_1,a_2,...,a_{N+1}=\Omega^*_1,\Omega^*_2,...,\Omega^*_N,\Delta \Omega$.

% \begin{equation}
% F(\mathrm{r}_j) =\left\{ 
%      \begin{array}{ll}
%          \Omega_\mathrm{1} &  0 < \mathrm{r/R}_\mathrm{1} \le \mathrm{r}_\mathrm{BCZ_1}\\     
%          \Omega_\mathrm{1} + \Delta \Omega_{*} & \mathrm{r}_\mathrm{BCZ_1} < \mathrm{r/R}_\mathrm{1} \le 1\\
%          ...\\
%          \Omega_\mathrm{N} & 0 < \mathrm{r/R}_N \le \mathrm{r}_\mathrm{BCZ_N}\\     
%          \Omega_\mathrm{N} + \Delta \Omega_{*} & \mathrm{r}_\mathrm{BCZ} < \mathrm{r/R}_N \le 
%      \end{array} \right.
% \end{equation}

%The splitting for the multiplet $i$ is then
%\begin{equation}
%\delta \omega_i = \sum_{j,k} K_{i}(r_j)F_{k}(r_j)a_k,
%\end{equation}
%where $K_i(r_j)$ are the kernels.
%A new matrix is formed $\mathbf{A}_{ik}=\sum_j \sigma_i(\delta \omega)^{-1} K_{ij} \phi_j F_{jk}$ which operates directly on $a_k$. 
%Then in matrix notation,
%\begin{equation}
%\mathbf{a}=(\mathbf{A}^T \mathbf{A})^{-1} \mathbf{A}^T \frac{\delta \omega_i}{\sigma(\delta \omega_i)}. 
%\end{equation}

\section{Independent asteroseismic fits}\label{sect:indfit}

Paper~I shows that inversions for stars using different stellar models within the spectroscopic uncertainties  have similar uncertainties.
To determine over which range of stars this property extends and thus over what range of stellar types an ensemble fit would be beneficial, we compute independent asteroseismic fits for each star.
First, we  compute the fit for each stellar model independently ($N=1$) using only $\ell=1$ modes. Table~\ref{table:invmodels} (Columns 4 and 5) shows the inverted rotation coefficients and uncertainties for each of the models. We impose  a different random noise realisation on the splittings for each star.
The uncertainties in the inverted rotation coefficients are large and, although the synthetic rotation rate below the BCZ and the step size are identical for each model, 
the inversion returns a broad range of $\Omega^*$ and ${\Delta \Omega}$ values, where the step can be negative or positive. 
If we double the number of splittings to use $\ell=1, |m|=1$ and $\ell=2, |m|=2$ modes (since we are using $i=90^\circ$), the uncertainties are reduced (Table~\ref{table:invmodels2}, Columns 4 and 5) but still not enough to successfully constrain the size or sign of the RDR gradient.
As shown in Paper~I  for regularised least squares inversions, we see that the uncertainties for each of the stellar models across the main sequence are of the same order.
This suggests that the uncertainty in the inverted coefficients will reduce as the square root of the number of stars, and all stars across the main sequence can be included  in an ensemble inversion.

%%%%%%%         RESULTS         %%%%%%%%%%
\section{Ensemble asteroseismic fits}\label{sect:ens}

We now explore the ensemble fit for all of the 15 stars ($N$) at once.
The ensemble fits of the $\ell=1$ modes for all of our modelled stars are shown in Table~\ref{table:invmodels}, Columns 6 and 7. One thing to note is that the weighted mean of the step size for the independent fits is $-216 \pm 294$~nHz, showing that the ensemble inversion  provides  slightly better uncertainties than simply averaging the independent inversions.
% make_textable_inversion_pmunc_inc.pro

The uncertainties are reduced from the independent cases and  the inverted values for $\Omega^*_N$ vary less, but the uncertainty is not small enough to constrain the sign of $\Delta \Omega$.
Using only the $\ell=1$ mode splittings, this method cannot differentiate between a solid body rotation and a rotation gradient for this synthetic profile. 

In Table~\ref{table:invmodels2} Columns 6 and 7, we show the results from fits including the $\ell=2, |m|=2$ mode splittings.  
The uncertainties in the ensemble fit show that the sign of the step can be correctly constrained since $\sigma(\Omega^*_N) < |\Delta \Omega_\mathrm{c}|$. 
For this noise realisation, the inverted values for $\Omega^*_N$ and ${\Delta \Omega}_{*}$ are quite close to the input values ($\Omega_\mathrm{c}=1035$~nHz and $\Delta \Omega_\mathrm{c}=-350$~nHz). 

\onecolumn

\begin{table}
\caption{Inferred rotation coefficients and uncertainties for $\ell=1, |m|=1$ mode splittings with a rotation axis inclination of  $90^\circ$ for independent fits to each star (Columns 4 and 5);  for an ensemble fit for stars with similar synthetic rotation profiles (Columns~6 and 7);  for an ensemble fit for stars with rotation profiles where the rotation rate below the BCZ varied randomly up to 50\% from $1035$~nHz (Columns~8 and 9).  }  % title of Table
\vspace{-0.5cm}
\label{table:invmodels} % is used to refer this table in the text
\centering % used for centering table     
\include{inv_mdstep_params_models_ens_ell1_90deg}   
% make_textable_inversion_pmunc_inc.pro
% invert_ensemble_step_inclination.pro
%\tablefoot{some comments here}
\end{table}

\begin{table}
\caption{Inferred rotation coefficients and uncertainties, as in Table~\ref{table:invmodels}, but now including $\ell=1, |m|=1$ and $\ell=2, |m|=1,2$ mode splittings with a rotation axis inclination of  $90^\circ$. }  % title of Table
\vspace{-0.5cm}
\label{table:invmodels2} % is used to refer this table in the text
\centering % used for centering table   
\include{inv_mdstep_params_models_ens_ell2_90deg}   
% make_textable_inversion_pmunc_inc.pro
% invert_ensemble_step_inclination.pro
%\tablefoot{some comments here}
\end{table}

\twocolumn

%%%%%%%%%%%%%%%%%%%%%%%%%%%%%%%%%%

\subsection{Variations in the rotation profile}\label{sect:vary}
This is a linear inversion and we can only invert for the average step size.
To demonstrate the stability of the method, we  compute the ensemble fit again, but with the synthetic rotation rate in the core  perturbed by a Gaussian distributed random realisation of the noise centred at zero and standard deviation of 50\% of $1035$~nHz ($\approx 250$~nHz)  (Tables~\ref{table:invmodels} and \ref{table:invmodels2}, Columns 8 and 9). The variation in values of $\Omega_\mathrm{c}$ is larger than when the synthetic $\Omega_\mathrm{c}$ are all the same, reflecting the variation of the synthetic rotation profile in the stars.
Varying the rotation rate below the convection zone was a choice to simply demonstrate the behaviour of the fitting method in response to an ensemble of different rotation rates. We could have equally  chosen to vary the step size, or both parameters. However, the resulting uncertainties in the parameters would have remained the same.

\subsection{Lower radial order mode splittings}\label{sect:lowfreq}

We explore the consequences for the inversions  by adding the next five lower  radial orders to all of the stars, so that we use a total of 30 mode splittings.
In Table~\ref{table:invlown}, we show   that the uncertainties are reduced, which demonstrates that splittings measured from lower frequency  modes better constrain the fit (Fig.~\ref{fig:ensnoise}).

\begin{table}
\caption{Results for inversions using  $\ell=1, |m|=1$ and $\ell=2, |m|=1,2$ mode splittings with a rotation axis inclination of  $90^\circ$, extending the radial order to the next five lowest radial orders  in each star.}  % title of Table
\vspace{-0.5cm}
\label{table:invlown} % is used to refer this table in the text
\centering % used for centering table    
\include{inv_mdstep_params_models_ens_ell2_lown_90deg}   
% make_textable_inversion_ens_lown_90deg.pro
% invert_ensemble_step_inclination.pro   with nlow=9 (not 4)
\end{table}

\section{Measuring radial differential rotation along the main sequence}\label{sect:plato}
Our ensemble fit successfully inferred the sign of the rotation rate for  a broad range of stellar models along the main sequence.
We now select three ensembles of stars based on spectral type to determine the sensitivity of our method to differences in rotation across the main sequence. This was simply a choice to demonstrate the potential of the ensemble fitting method, however it is not known how or if differential rotation varies systematically across the main sequence. This is something we would like to determine in the future using our ensemble fitting method. Analysing ensembles of stars based on other physical characteristics, such as the bulk rotation rate, activity level, or age, are other options.

We estimate  the uncertainty in the ensemble fits using 1000 stars.
We estimate the proportion of stellar types using the stellar counts from the Geneva-Copenhagen Survey \citep{Nordstrometal2004} of bright, main-sequence cool dwarfs. 
Using the temperatures, we estimate that approximately 
60\% will be F-type stars ($T_\mathrm{eff} \ge 6000$~K), 
35\% will be G-type ($5000 \le T_\mathrm{eff}<6000$~K) stars, and 
5\% K-type ($T_\mathrm{eff}<5000$~K) stars. 
We note that PLATO will make two long duration observations of  $>1000$ bright stars (as in the Geneva-Copenhagen Survey) of stellar type F5 to K7  and 20,000 less bright stars, that will be suitable for measuring rotation \citep{yellowbook}.

As shown by Paper~I and in Sect.~\ref{sect:indfit} of this paper, the uncertainties for all the independent inversions for each star will reduce with the square root of the number of stars used in the ensemble fit. Table~\ref{table:plato} shows the mean of the uncertainties for all the models of each stellar type in our sample.
If we extrapolate these uncertainties to the number of stars of each type expected to be observed by PLATO, the uncertainties can be significantly reduced (right-hand side of Table~\ref{table:plato} ).
Therefore, given the uncertainties on the step, $\sigma({\Delta \Omega}_\mathrm{*})$, in Table~\ref{table:plato}, it would be possible to distinguish between the average RDR gradient for different stellar types across the main sequence if the differences are larger than $\approx 200$~nHz.

\begin{table}
\caption{Estimated uncertainties on the inverted rotation coefficients for the 15 stars of different types in our sample (top) and 1000 stars (bottom) for only $\ell=1$ modes, with the standard 10 radial orders centred about $\nu_\mathrm{max}$.}
\include{plato_types_estimate}
\tablefoot{Instead of listing the uncertainty for each type of stellar model, $\sigma(\Omega^*_N)$, we list the mean of the uncertainties, $\langle \sigma(\Omega^*_N) \rangle$.}
 \label{table:plato}
 \end{table}
%   invert_ensemble_types_step_inc.pro
%   make_plato_table.pro
%    38./sqrt(nstars/15.)     58/sqrt(nstars/15.)       , where nstars is 1800,  800 or 100

Thus far, we have only used uncertainties for stars with a rotation axis perpendicular to the line of sight ($i=90^\circ$). Assuming stars have a rotation axis randomly inclined to the line of sight, the  probability distribution of a star having inclination $i$ is a $\sin i$  distribution. To estimate the effect of stellar inclination on the uncertainties on the fitting coefficients, we computed the functional fit for each star with inclinations from $10^\circ$ to $90^\circ$ and computed the $\sin i$ weighted average. The uncertainty on the ensemble of each type of star in our sample is shown in Table~\ref{table:platoinc} (left). On the right, we have  scaled for the number of stars in a sample of 1000 (Table~\ref{table:platoinc}, right). 
The uncertainties on the fits are larger at lower inclinations and so there is a slightly larger uncertainty than at $i=90^\circ$.The smallest uncertainties that are estimated for  on the step, $\sigma({\Delta \Omega}_\mathrm{*})$, in Table~\ref{table:platoinc} for $\ell=1,2$ modes show that it would be possible to distinguish between the average RDR gradient for different stellar types across the main sequence if the differences are larger than $\approx 100$~nHz. The uncertainties on the RDR themselves allow us to place constraints down to the order of solar RDR ($\approx 30$~nHz). The trend in the rotation rate could be constrained if the differences are larger than $\approx 50$~nHz.

\begin{table}
\caption{Estimated uncertainties weighted by the distribution of expected inclinations on the inverted rotation coefficients.}
\include{plato_types_estimate_inc}
\tablefoot{Stars of each main-sequence type in our sample of stellar models (left) and 1000 stars (right). Using only $\ell=1$ modes in the top half and $\ell=1,2$ modes in the lower half of the table with the standard 10 radial orders centred about $\nu_\mathrm{max}$.}

 \label{table:platoinc}
 %   invert_ensemble_types_step_inc.pro
 % compute_ensemble_uncertainties_inclination.pro
 %   make_plato_table.pro
 \end{table}

%%%%%%%%%%%%%%%%%
 \section{Conclusion}\label{sect:conc}
 
Supported by the fact that the inversions are insensitive to the uncertainties in the stellar models  (Paper~I), we found that ensemble asteroseismic fits are a feasible method to extract basic rotation parameters. 
We found that the uncertainties in the independent RDR fits of F, G, and K-type stars are of the same order (Sect.~\ref{sect:indfit}).
We demonstrated that for 15 stars across the main sequence it is possible to distinguish between solid body rotation and a RDR gradient of $\approx 200$~nHz using splittings of $\ell=1,2$ modes.

However, ensemble fittings and inversions have limitations. 
The assumption behind ensemble fitting is that the RDR of the stars in the ensemble behaves systematically. 
This technique is a linear inversion to retrieve the average RDR step function size. 
Any outliers will have an effect on the result depending on the number of modes used in the inversion, the level of noise on the splittings, and the magnitude of the outlier. 
For example, if by some coincidence half of the ensemble stars had a RDR with exactly the opposite sign and magnitude of the other half, then the retrieved RDR from the ensemble fitting would indicate solid-body rotation. 
Although, the variance of the inferred values for the rotation below the convection zone does reflect the variance of the input values.
Since the RDR of stars along the main sequence is not known, and this is what we would like to constrain, there is no prior information to help select a useful ensemble. Therefore, doing ensemble fitting for ensembles of stars based on different physical characteristics, such as bulk rotation rate, activity level, or age, will help us to understand for which physical processes RDR is important.

Further constraints to the inversions for RDR could be implemented by using  surface rotation constraints from starspot rotation  or $v \sin i$ measurements.
Measuring rotational splittings of  lower radial order or higher harmonic degree rotational splittings with SONG \citep[][, a new ground-based network to measure the surface Doppler velocities of stars]{Palleetal2013} may also help to reduce the uncertainties, especially at lower frequencies.

With the advent of PLATO, we will be able to observe thousands of main-sequence Sun-like stars, and begin to trace differences in rotation for subgroups of stellar populations: e.g. stellar type, bulk rotation rate, or activity-related measurements. 
We showed that with a total of 1000 stars, the difference in magnitude of the gradient  can be detected for F, G, and K-type stars if the difference in the average RDR gradient  is larger than $\approx 100$~nHz. We also showed that this would place constraints on the average RDR of the ensemble of stars down to the order of solar RDR, and we could detect trends in the rotation rate across the main sequence above $\approx 50$~nHz.
Ensemble inversions can also help to constrain the importance of RDR for stellar activity such as  absolute activity level,  stellar cycle period, and spot coverage.

%%%%%%%%%%%%%%%%%
\begin{acknowledgements}
The authors acknowledge research funding by Deutsche Forschungsgemeinschaft (DFG) under grant SFB 963/1 ``Astrophysical flow instabilities and turbulence'' (Project A18). 
\end{acknowledgements}

\bibliographystyle{aa} % style aa.bst
\bibliography{rotinv} % your references Yourfile.bib

\end{document}

%% file: inv_mdstep_params_models_ens_ell1_90deg.tex
\begin{tabular}{ l  c c || c c  || c c | c c  }
 \,   & \,   & \,  & \multicolumn{2}{c}{independent inversions} &  \multicolumn{4}{c}{ensemble inversions}   \\
 \,   & \,   & \,  & \, & \,  & \,  & \, & \multicolumn{2}{c}{$\Omega_c(1\pm0.5)$}  \\
  T$_\mathrm{eff}$ & $r_\mathrm{BCZ}$  &  $\langle n\rangle$   &  $\Omega^* \pm \sigma(\Omega^*) $    &  $\Delta \Omega \pm \sigma(\Delta \Omega)$ &      $\Omega^*_N \pm \sigma(\Omega^*_N)$    & $\Delta \Omega \pm \sigma(\Delta \Omega)$    &         $\Omega^*_N \pm \sigma(\Omega^*_N)$ &  $\Delta \Omega \pm \sigma(\Delta \Omega)$   \\
  $[K]$  & [$R$] & \, &  [nHz] & [nHz] & [nHz] & [nHz] &  [nHz] & [nHz]    \\
\hline \\
4538 & 0.65  &  27 & 693 $\pm$ 552 & 828 $\pm$ 1073 & 883 $\pm$ 173 & -111 $\pm$ 266  &  449 $\pm$ 173 & -111 $\pm$ 266 \\ 
4676 & 0.64  &  24 & 1810 $\pm$ 607 & -838 $\pm$ 720 & 874 $\pm$ 178 & -111 $\pm$ 266  &  441 $\pm$ 178 & -111 $\pm$ 266 \\ 
4860 & 0.61  &  22 & 1632 $\pm$ 676 & -641 $\pm$ 797 & 866 $\pm$ 184 & -111 $\pm$ 266  &  324 $\pm$ 184 & -111 $\pm$ 266 \\ 
5599 & 0.73  &  22 & 1828 $\pm$ 587 & -1360 $\pm$ 1146 & 876 $\pm$ 164 & -111 $\pm$ 266  &  1146 $\pm$ 164 & -111 $\pm$ 266 \\ 
5627 & 0.67  &  20 & 952 $\pm$ 699 & 254 $\pm$ 1336 & 890 $\pm$ 166 & -111 $\pm$ 266  &  726 $\pm$ 166 & -111 $\pm$ 266 \\ 
5632 & 0.71  &  23 & 1422 $\pm$ 708 & -625 $\pm$ 1378 & 865 $\pm$ 174 & -111 $\pm$ 266  &  904 $\pm$ 174 & -111 $\pm$ 266 \\ 
5701 & 0.70  &  22 & -467 $\pm$ 723 & 3056 $\pm$ 1394 & 925 $\pm$ 126 & -111 $\pm$ 266  &  795 $\pm$ 126 & -111 $\pm$ 266 \\ 
5744 & 0.72  &  20 & -230 $\pm$ 627 & 2640 $\pm$ 1215 & 923 $\pm$ 137 & -111 $\pm$ 266  &  1421 $\pm$ 137 & -111 $\pm$ 266 \\ 
5892 & 0.71  &  18 & 1823 $\pm$ 620 & -1358 $\pm$ 1195 & 880 $\pm$ 156 & -111 $\pm$ 266  &  718 $\pm$ 156 & -111 $\pm$ 266 \\ 
5983 & 0.75  &  19 & 1649 $\pm$ 723 & -1078 $\pm$ 1405 & 891 $\pm$ 158 & -111 $\pm$ 266  &  961 $\pm$ 158 & -111 $\pm$ 266 \\ 
6252 & 0.81  &  19 & 1382 $\pm$ 592 & -464 $\pm$ 1181 & 890 $\pm$ 162 & -111 $\pm$ 266  &  791 $\pm$ 162 & -111 $\pm$ 266 \\ 
6408 & 0.84  &  21 & 1019 $\pm$ 583 & 271 $\pm$ 1170 & 890 $\pm$ 167 & -111 $\pm$ 266  &  1733 $\pm$ 167 & -111 $\pm$ 266 \\ 
6661 & 0.91  &  17 & 942 $\pm$ 422 & 743 $\pm$ 1392 & 949 $\pm$ 109 & -111 $\pm$ 266  &  493 $\pm$ 109 & -111 $\pm$ 266 \\ 
6663 & 0.88  &  19 & 380 $\pm$ 472 & 2591 $\pm$ 1518 & 931 $\pm$ 102 & -111 $\pm$ 266  &  1120 $\pm$ 102 & -111 $\pm$ 266 \\ 
6680 & 0.89  &  18 & 1121 $\pm$ 431 & 112 $\pm$ 1407 & 942 $\pm$ 97 & -111 $\pm$ 266  &  1123 $\pm$ 97 & -111 $\pm$ 266 \\ 
\end{tabular}

%% file: inv_mdstep_params_models_ens_ell2_90deg.tex
\begin{tabular}{ l  c c || c c  || c c | c c  }
 \,   & \,   & \,  & \multicolumn{2}{c}{independent inversions} &  \multicolumn{4}{c}{ensemble inversions}   \\
 \,   & \,   & \,  & \, & \,  & \,  & \, & \multicolumn{2}{c}{$\Omega_c(1\pm0.5)$}  \\
  T$_\mathrm{eff}$ & $r_\mathrm{BCZ}$  &  $\langle n\rangle$   &  $\Omega^* \pm \sigma(\Omega^*) $    &  $\Delta \Omega \pm \sigma(\Delta \Omega)$ &      $\Omega^*_N \pm \sigma(\Omega^*_N)$    & $\Delta \Omega \pm \sigma(\Delta \Omega)$    &         $\Omega^*_N \pm \sigma(\Omega^*_N)$ &  $\Delta \Omega \pm \sigma(\Delta \Omega)$   \\
  $[K]$  & [$R$] & \, &  [nHz] & [nHz] & [nHz] & [nHz] &  [nHz] & [nHz]    \\
\hline \\
4538 & 0.65  &  27 & 1015 $\pm$ 111 & 181 $\pm$ 217 & 1024 $\pm$ 49 & -332 $\pm$ 75  &  591 $\pm$ 49 & -332 $\pm$ 75 \\ 
4676 & 0.64  &  24 & 948 $\pm$ 161 & 185 $\pm$ 191 & 1024 $\pm$ 50 & -332 $\pm$ 75  &  591 $\pm$ 50 & -332 $\pm$ 75 \\ 
4860 & 0.61  &  22 & -32 $\pm$ 231 & 1317 $\pm$ 272 & 1022 $\pm$ 52 & -332 $\pm$ 75  &  479 $\pm$ 52 & -332 $\pm$ 75 \\ 
5599 & 0.73  &  22 & 1396 $\pm$ 188 & -539 $\pm$ 368 & 1021 $\pm$ 46 & -332 $\pm$ 75  &  1290 $\pm$ 46 & -332 $\pm$ 75 \\ 
5627 & 0.67  &  20 & 1205 $\pm$ 170 & -229 $\pm$ 329 & 1019 $\pm$ 47 & -332 $\pm$ 75  &  855 $\pm$ 47 & -332 $\pm$ 75 \\ 
5632 & 0.71  &  23 & 1292 $\pm$ 167 & -352 $\pm$ 326 & 1022 $\pm$ 49 & -332 $\pm$ 75  &  1062 $\pm$ 49 & -332 $\pm$ 75 \\ 
5701 & 0.70  &  22 & 1182 $\pm$ 162 & -163 $\pm$ 314 & 1026 $\pm$ 35 & -332 $\pm$ 75  &  896 $\pm$ 35 & -332 $\pm$ 75 \\ 
5744 & 0.72  &  20 & 943 $\pm$ 162 & 326 $\pm$ 316 & 1026 $\pm$ 38 & -332 $\pm$ 75  &  1524 $\pm$ 38 & -332 $\pm$ 75 \\ 
5892 & 0.71  &  18 & 1076 $\pm$ 160 & 44 $\pm$ 310 & 1019 $\pm$ 44 & -332 $\pm$ 75  &  857 $\pm$ 44 & -332 $\pm$ 75 \\ 
5983 & 0.75  &  19 & 1218 $\pm$ 209 & -236 $\pm$ 406 & 1020 $\pm$ 45 & -332 $\pm$ 75  &  1090 $\pm$ 45 & -332 $\pm$ 75 \\ 
6252 & 0.81  &  19 & 1390 $\pm$ 403 & -523 $\pm$ 807 & 1019 $\pm$ 45 & -332 $\pm$ 75  &  919 $\pm$ 45 & -332 $\pm$ 75 \\ 
6408 & 0.84  &  21 & 1101 $\pm$ 318 & 75 $\pm$ 643 & 1022 $\pm$ 47 & -332 $\pm$ 75  &  1865 $\pm$ 47 & -332 $\pm$ 75 \\ 
6661 & 0.91  &  16 & 1283 $\pm$ 134 & -349 $\pm$ 438 & 1028 $\pm$ 31 & -332 $\pm$ 75  &  571 $\pm$ 31 & -332 $\pm$ 75 \\ 
6663 & 0.88  &  19 & 1424 $\pm$ 164 & -830 $\pm$ 531 & 1020 $\pm$ 29 & -332 $\pm$ 75  &  1209 $\pm$ 29 & -332 $\pm$ 75 \\ 
6680 & 0.89  &  18 & 1233 $\pm$ 132 & -224 $\pm$ 430 & 1025 $\pm$ 27 & -332 $\pm$ 75  &  1206 $\pm$ 27 & -332 $\pm$ 75 \\ 
\end{tabular}

%% file: inv_mdstep_params_models_ens_ell2_lown_90deg.tex
\begin{tabular}{ l  c c | c c   }
  T$_\mathrm{eff}$ & $r_\mathrm{BCZ}$  &  $\langle n\rangle$  &    $\Omega^*_N \pm \sigma(\Omega^*_N)$    & $\Delta \Omega \pm \sigma(\Delta \Omega)$    \\ 
  $[K]$  & [$R$] & \, &  [nHz] & [nHz]   \\
\hline \\
4538 & 0.65  &  24 & 1026 $\pm$ 21 & -337 $\pm$ 32 \\ 
4676 & 0.64  &  21 & 1026 $\pm$ 22 & -337 $\pm$ 32 \\ 
4860 & 0.61  &  19 & 1026 $\pm$ 22 & -337 $\pm$ 32 \\ 
5599 & 0.73  &  19 & 1027 $\pm$ 20 & -337 $\pm$ 32 \\ 
5627 & 0.67  &  17 & 1025 $\pm$ 20 & -337 $\pm$ 32 \\ 
5632 & 0.71  &  20 & 1025 $\pm$ 22 & -337 $\pm$ 32 \\ 
5701 & 0.70  &  19 & 1028 $\pm$ 15 & -337 $\pm$ 32 \\ 
5744 & 0.72  &  17 & 1028 $\pm$ 17 & -337 $\pm$ 32 \\ 
5892 & 0.71  &  15 & 1025 $\pm$ 19 & -337 $\pm$ 32 \\ 
5983 & 0.75  &  16 & 1026 $\pm$ 19 & -337 $\pm$ 32 \\ 
6252 & 0.81  &  16 & 1026 $\pm$ 20 & -337 $\pm$ 32 \\ 
6408 & 0.84  &  18 & 1024 $\pm$ 20 & -337 $\pm$ 32 \\ 
6661 & 0.91  &  13 & 1028 $\pm$ 13 & -337 $\pm$ 32 \\ 
6663 & 0.88  &  16 & 1026 $\pm$ 12 & -337 $\pm$ 32 \\ 
6680 & 0.89  &  15 & 1027 $\pm$ 11 & -337 $\pm$ 32 \\ 
\end{tabular}

%% file: plato_types_estimate.tex
\begin{tabular}{c  c | c c || c | c c }
  Stellar & $N_\mathrm{stars}$   & $\langle \sigma(\Omega^*_N) \rangle$ & $\sigma(\Delta \Omega) $ & $N_\mathrm{stars}$   & $\langle \sigma(\Omega^*_N) \rangle$ & $\sigma(\Delta \Omega) $\\
   Type &   & [nHz] & [nHz] &  & [nHz] & [nHz] \\
\hline 
  F &  5 &  226 &  527 &  600 &  20 &  48 \\
  G &  7 &  245 &  398 &  350 &  34 &  56 \\
  K &  3 &  327 &  488 &   50 &  80 & 119 \\ 
 \end{tabular}

%% file: plato_types_estimate_inc.tex
\begin{tabular}{c  c | c c || c | c c}
  Stellar & $N_\mathrm{stars}$   & $\langle \sigma(\Omega^*_N) \rangle$ & $\sigma(\Delta \Omega) $ & $N_\mathrm{stars}$   & $\langle \sigma(\Omega^*_N) \rangle$ & $\sigma(\Delta \Omega) $\\
   Type  &   & [nHz] & [nHz] &   & [nHz] & [nHz] \\
\hline  
  F &  5 &  232 &  628 &  600 &   21 &   57 \\
  G &  7 &  262 &  508 &  350 &   37 &   71 \\
  K &  3 &  357 &  489 &   50 &   87 &  119 \\
\hline \hline  
  F &  5 &   90 &  273 &  600 &    8 &   24 \\
  G &  7 &   72 &  140 &  350 &   10 &   19 \\
  K &  3 &   88 &  133 &   50 &   21 &   32 \\
 \end{tabular}

%% file: ms_final.bbl
\begin{thebibliography}{31}
\expandafter\ifx\csname natexlab\endcsname\relax\def\natexlab#1{#1}\fi

\bibitem[{{Angulo} {et~al.}(1999){Angulo}, {Arnould}, {Rayet}, {Descouvemont},
  {Baye}, {Leclercq-Willain}, {Coc}, {Barhoumi}, {Aguer}, {Rolfs}, {Kunz},
  {Hammer}, {Mayer}, {Paradellis}, {Kossionides}, {Chronidou}, {Spyrou},
  {degl'Innocenti}, {Fiorentini}, {Ricci}, {Zavatarelli}, {Providencia},
  {Wolters}, {Soares}, {Grama}, {Rahighi}, {Shotter}, \& {Lamehi
  Rachti}}]{Angulo1999}
{Angulo}, C., {Arnould}, M., {Rayet}, M., {et~al.} 1999, Nuclear Physics A,
  656, 3

\bibitem[{{B{\"o}hm-Vitense}(1958)}]{Boehm-vitense1958}
{B{\"o}hm-Vitense}, E. 1958, \zap, 46, 108

\bibitem[{{Borucki} {et~al.}(2010){Borucki}, {Koch}, {Basri}, {Batalha},
  {Brown}, {Caldwell}, {Caldwell}, {Christensen-Dalsgaard}, {Cochran},
  {DeVore}, {Dunham}, {Dupree}, {Gautier}, {Geary}, {Gilliland}, {Gould},
  {Howell}, {Jenkins}, {Kondo}, {Latham}, {Marcy}, {Meibom}, {Kjeldsen},
  {Lissauer}, {Monet}, {Morrison}, {Sasselov}, {Tarter}, {Boss}, {Brownlee},
  {Owen}, {Buzasi}, {Charbonneau}, {Doyle}, {Fortney}, {Ford}, {Holman},
  {Seager}, {Steffen}, {Welsh}, {Rowe}, {Anderson}, {Buchhave}, {Ciardi},
  {Walkowicz}, {Sherry}, {Horch}, {Isaacson}, {Everett}, {Fischer}, {Torres},
  {Johnson}, {Endl}, {MacQueen}, {Bryson}, {Dotson}, {Haas}, {Kolodziejczak},
  {Van Cleve}, {Chandrasekaran}, {Twicken}, {Quintana}, {Clarke}, {Allen},
  {Li}, {Wu}, {Tenenbaum}, {Verner}, {Bruhweiler}, {Barnes}, \&
  {Prsa}}]{Boruckietal2010}
{Borucki}, W.~J., {Koch}, D., {Basri}, G., {et~al.} 2010, Science, 327, 977

\bibitem[{{Caughlan} \& {Fowler}(1988)}]{Caughlan1988}
{Caughlan}, G.~R. \& {Fowler}, W.~A. 1988, Atomic Data and Nuclear Data Tables,
  40, 283

\bibitem[{{Chaplin} {et~al.}(2009){Chaplin}, {Houdek}, {Karoff}, {Elsworth}, \&
  {New}}]{Chaplinetal2009}
{Chaplin}, W.~J., {Houdek}, G., {Karoff}, C., {Elsworth}, Y., \& {New}, R.
  2009, \aap, 500, L21

\bibitem[{{Charbonneau}(2010)}]{Charbonneau2010}
{Charbonneau}, P. 2010, Living Reviews in Solar Physics, 7, 3

\bibitem[{{Deheuvels} {et~al.}(2014){Deheuvels}, {Do{\u g}an}, {Goupil},
  {Appourchaux}, {Benomar}, {Bruntt}, {Campante}, {Casagrande}, {Ceillier},
  {Davies}, {De Cat}, {Fu}, {Garc{\'{\i}}a}, {Lobel}, {Mosser}, {Reese},
  {Regulo}, {Schou}, {Stahn}, {Thygesen}, {Yang}, {Chaplin},
  {Christensen-Dalsgaard}, {Eggenberger}, {Gizon}, {Mathis},
  {Molenda-{\.Z}akowicz}, \& {Pinsonneault}}]{Deheuvelsetal2014}
{Deheuvels}, S., {Do{\u g}an}, G., {Goupil}, M.~J., {et~al.} 2014, \aap, 564,
  A27

\bibitem[{{Deheuvels} {et~al.}(2012){Deheuvels}, {Garc{\'{\i}}a}, {Chaplin},
  {Basu}, {Antia}, {Appourchaux}, {Benomar}, {Davies}, {Elsworth}, {Gizon},
  {Goupil}, {Reese}, {Regulo}, {Schou}, {Stahn}, {Casagrande},
  {Christensen-Dalsgaard}, {Fischer}, {Hekker}, {Kjeldsen}, {Mathur}, {Mosser},
  {Pinsonneault}, {Valenti}, {Christiansen}, {Kinemuchi}, \&
  {Mullally}}]{Deheuvelsetal2012}
{Deheuvels}, S., {Garc{\'{\i}}a}, R.~A., {Chaplin}, W.~J., {et~al.} 2012, \apj,
  756, 19

\bibitem[{{Ferguson} {et~al.}(2005){Ferguson}, {Alexander}, {Allard}, {Barman},
  {Bodnarik}, {Hauschildt}, {Heffner-Wong}, \& {Tamanai}}]{Ferguson2005}
{Ferguson}, J.~W., {Alexander}, D.~R., {Allard}, F., {et~al.} 2005, \apj, 623,
  585

\bibitem[{Gizon {et~al.}(2013)Gizon, Ballot, Michel, Stahn, Vauclair, Bruntt,
  Quirion, Benomar, Vauclair, Appourchaux, Auvergne, Baglin, Barban, Baudin,
  Bazot, Campante, Catala, Chaplin, Creevey, Deheuvels, Dolez, Elsworth,
  Garcia, Gaulme, Mathis, Mathur, Mosser, Regulo, Roxburgh, Salabert, Samadi,
  Sato, Verner, Hanasoge, \& Sreenivasan}]{Gizonetal2013}
Gizon, L., Ballot, J., Michel, E., {et~al.} 2013, Proceedings of the National
  Academy of Sciences, 110, 13267

\bibitem[{Gizon \& Solanki(2004)}]{GizonSolanki2004}
Gizon, L. \& Solanki, S.~K. 2004, Solar Physics, 220, 169

\bibitem[{{Grevesse} \& {Sauval}(1998)}]{Grevesse1998}
{Grevesse}, N. \& {Sauval}, A.~J. 1998, \ssr, 85, 161

\bibitem[{{Harvey}(1985)}]{Harvey1985}
{Harvey}, J. 1985, in ESA Special Publication, Vol. 235, Future Missions in
  Solar, Heliospheric \& Space Plasma Physics, ed. E.~{Rolfe} \& B.~{Battrick},
  199--208

\bibitem[{{Iglesias} \& {Rogers}(1996)}]{Iglesias1996}
{Iglesias}, C.~A. \& {Rogers}, F.~J. 1996, \apj, 464, 943

\bibitem[{{Imbriani} {et~al.}(2005){Imbriani}, {Costantini}, {Formicola},
  {Vomiero}, {Angulo}, {Bemmerer}, {Bonetti}, {Broggini}, {Confortola},
  {Corvisiero}, {Cruz}, {Descouvemont}, {F{\"u}l{\"o}p}, {Gervino},
  {Guglielmetti}, {Gustavino}, {Gy{\"u}rky}, {Jesus}, {Junker}, {Klug},
  {Lemut}, {Menegazzo}, {Prati}, {Roca}, {Rolfs}, {Romano}, {Rossi-Alvarez},
  {Sch{\"u}mann}, {Sch{\"u}rmann}, {Somorjai}, {Straniero}, {Strieder},
  {Terrasi}, \& {Trautvetter}}]{Imbriani2005}
{Imbriani}, G., {Costantini}, H., {Formicola}, A., {et~al.} 2005, European
  Physical Journal A, 25, 455

\bibitem[{{Kunz} {et~al.}(2002){Kunz}, {Fey}, {Jaeger}, {Mayer}, {Hammer},
  {Staudt}, {Harissopulos}, \& {Paradellis}}]{Kunz2002}
{Kunz}, R., {Fey}, M., {Jaeger}, M., {et~al.} 2002, \apj, 567, 643

\bibitem[{{Libbrecht}(1992)}]{Libbrecht1992}
{Libbrecht}, K.~G. 1992, \apj, 387, 712

\bibitem[{{Lund} {et~al.}(2014){Lund}, {Miesch}, \&
  {Christensen-Dalsgaard}}]{Lundetal2014b}
{Lund}, M.~N., {Miesch}, M.~S., \& {Christensen-Dalsgaard}, J. 2014, \apj, 790,
  121

\bibitem[{{Nielsen} {et~al.}(2014){Nielsen}, {Gizon}, {Schunker}, \&
  {Schou}}]{Nielsenetal2014}
{Nielsen}, M., {Gizon}, L., {Schunker}, H., \& {Schou}, J. 2014, \aap

\bibitem[{{Nielsen} {et~al.}(2015){Nielsen}, {Schunker}, {Gizon}, \&
  {Ball}}]{Nielsenetal2015}
{Nielsen}, M., {Schunker}, H., {Gizon}, L., \& {Ball}, W. 2015, \aap

\bibitem[{{Nordstr{\"o}m} {et~al.}(2004){Nordstr{\"o}m}, {Mayor}, {Andersen},
  {Holmberg}, {Pont}, {J{\o}rgensen}, {Olsen}, {Udry}, \&
  {Mowlavi}}]{Nordstrometal2004}
{Nordstr{\"o}m}, B., {Mayor}, M., {Andersen}, J., {et~al.} 2004, \aap, 418, 989

\bibitem[{{Pall{\'e}} {et~al.}(2013){Pall{\'e}}, {Grundahl}, {Trivi{\~n}o
  Hage}, {Christensen-Dalsgaard}, {Frandsen}, {Garc{\'{\i}}a}, {Uytterhoeven},
  {Andersen}, {Rasmussen}, {S{\o}rensen}, {Kjeldsen}, {Spano}, {Nilsson},
  {Hartman}, {J{\o}rgensen}, {Skottfelt}, {Harps{\o}e}, \&
  {Andersen}}]{Palleetal2013}
{Pall{\'e}}, P.~L., {Grundahl}, F., {Trivi{\~n}o Hage}, A., {et~al.} 2013,
  Journal of Physics Conference Series, 440, 012051

\bibitem[{{Paxton} {et~al.}(2013){Paxton}, {Cantiello}, {Arras}, {Bildsten},
  {Brown}, {Dotter}, {Mankovich}, {Montgomery}, {Stello}, {Timmes}, \&
  {Townsend}}]{Paxton2013}
{Paxton}, B., {Cantiello}, M., {Arras}, P., {et~al.} 2013, \apjs, 208, 4

\bibitem[{{Rauer} {et~al.}(2013{\natexlab{a}}){Rauer}, {Catala}, {Aerts},
  {Appourchaux}, {Benz}, {Brandeker}, {Christensen-Dalsgaard}, {Deleuil},
  {Gizon}, {G{\"u}del}, {Janot-Pacheco}, {Mas-Hesse}, {Pagano}, {Piotto},
  {Pollacco}, {Santos}, {Smith}, {-C.}, {Su{\'a}rez}, {Szab{\'o}}, {Udry},
  {Adibekyan}, {Alibert}, {Almenara}, {Amaro-Seoane}, {Ammler-von Eiff},
  {Antonello}, {Ball}, {Barnes}, {Baudin}, {Belkacem}, {Bergemann}, {Birch},
  {Boisse}, {Bonomo}, {Borsa}, {Brand{\~a}o}, {Brocato}, {Brun}, {Burleigh},
  {Burston}, {Cabrera}, {Cassisi}, {Chaplin}, {Charpinet}, {Chiappini},
  {Csizmadia}, {Cunha}, {Damasso}, {Davies}, {Deeg}, {de Oliveira Fialho},
  {Diaz}, {Dreizler}, {Dreyer}, {Eggenberger}, {Ehrenreich}, {Eigm{\"u}ller},
  {Erikson}, {Farmer}, {Feltzing}, {Figueira}, {Forveille}, {Fridlund},
  {Garc{\'{\i}}a}, {Giuffrida}, {Godolt}, {Gomes da Silva}, {Goupil},
  {Granzer}, {Grenfell}, {Grotsch-Noels}, {G{\"u}nther}, {Haswell}, {Hatzes},
  {H{\'e}brard}, {Hekker}, {Helled}, {Heng}, {Jenkins}, {Khodachenko},
  {Kislyakova}, {Kley}, {Kolb}, {Krivova}, {Kupka}, {Lammer}, {Lanza},
  {Lebreton}, {Magrin}, {Marcos-Arenal}, {Marrese}, {Marques}, {Martins},
  {Mathis}, {Mathur}, {Messina}, {Miglio}, {Montalban}, {Montalto}, {Monteiro},
  {Moradi}, {Moravveji}, {Mordasini}, {Morel}, {Mortier}, {Nascimbeni},
  {Nielsen}, {Noack}, {Norton}, {Ofir}, {Oshagh}, {Ouazzani}, {P{\'a}pics},
  {Parro}, {Petit}, {Plez}, {Poretti}, {Quirrenbach}, {Ragazzoni}, {Raimondo},
  {Rainer}, {Reese}, {Redmer}, {Reffert}, {Rojas-Ayala}, {Roxburgh}, {Solanki},
  {Salmon}, {Santerne}, {Schneider}, {Schou}, {Schuh}, {Schunker},
  {Silva-Valio}, {Silvotti}, {Skillen}, {Snellen}, {Sohl}, {Sousa}, {Sozzetti},
  {Stello}, {Strassmeier}, {{\v S}vanda}, {Szab{\'o}}, {Tkachenko}, {Valencia},
  {van Grootel}, {Vauclair}, {Ventura}, {Wagner}, {Walton}, {Weingrill},
  {Werner}, {Wheatley}, \& {Zwintz}}]{Rauer2013}
{Rauer}, H., {Catala}, C., {Aerts}, C., {et~al.} 2013{\natexlab{a}}, ArXiv
  e-prints

\bibitem[{{Rauer} {et~al.}(2013{\natexlab{b}}){Rauer}, {Pollaco}, {Goupil},
  {Giampaolo}, {Udry}, {Gizon}, {Gondoin}, {Piersanti}, {Heras}, {Fridlund},
  {Stankov}, {Baldesarra}, {O'Rourke}, {Parmar}, \& {The PLATO
  Team}}]{yellowbook}
{Rauer}, H., {Pollaco}, D., {Goupil}, M.-J., {et~al.} 2013{\natexlab{b}}, PLATO
  Assessment Study Report (Yellow Book) ESA/SRE(2013)5 (ESA)

\bibitem[{{Rogers} \& {Nayfonov}(2002)}]{Rogers2002}
{Rogers}, F.~J. \& {Nayfonov}, A. 2002, \apj, 576, 1064

\bibitem[{{Schou} {et~al.}(1998){Schou}, {Antia}, {Basu}, {Bogart}, {Bush},
  {Chitre}, {Christensen-Dalsgaard}, {Di Mauro}, {Dziembowski}, {Eff-Darwich},
  {Gough}, {Haber}, {Hoeksema}, {Howe}, {Korzennik}, {Kosovichev}, {Larsen},
  {Pijpers}, {Scherrer}, {Sekii}, {Tarbell}, {Title}, {Thompson}, \&
  {Toomre}}]{Schouetal1998}
{Schou}, J., {Antia}, H.~M., {Basu}, S., {et~al.} 1998, \apj, 505, 390

\bibitem[{{Schunker} {et~al.}(accepted){Schunker}, {Schou}, \&
  {Ball}}]{Schunkeretal2015a}
{Schunker}, H., {Schou}, J., \& {Ball}, W. accepted, \aap

\bibitem[{{Spiegel} \& {Zahn}(1992)}]{SpiegelZahn1992}
{Spiegel}, E.~A. \& {Zahn}, J.-P. 1992, \aap, 265, 106

\bibitem[{Stahn(2011)}]{Stahn2011}
Stahn, T. 2011, PhD thesis, Georg-August Universit\"at, G\"ottingen, Germany

\bibitem[{{Thoul} {et~al.}(1994){Thoul}, {Bahcall}, \& {Loeb}}]{Thoul1994}
{Thoul}, A.~A., {Bahcall}, J.~N., \& {Loeb}, A. 1994, \apj, 421, 828

\end{thebibliography}
